\documentclass[
prd
, twocolumn%
,showpacs ,amssymb,superscriptaddress,aps
]{revtex4}
\usepackage{graphicx}
\input{epsf}

\usepackage{amsmath,amssymb}
\usepackage{bm}

\newcommand{\dalm}{\kern1pt\vbox{\hrule height 0.9pt\hbox{\vrule width 0.9pt
\hskip 2.5pt\vbox{\vskip 5.5pt}\hskip 3pt\vrule width 0.3pt}\hrule height 0.3pt}
\kern1pt}

\begin{document}



\title{Constraining the density dependence of the nuclear symmetry energy from 
an X-ray bursting neutron star}

\author{Hajime Sotani}
\email{sotani@yukawa.kyoto-u.ac.jp}
\affiliation{Division of Theoretical Astronomy, National Astronomical Observatory of Japan, 2-21-1 Osawa, Mitaka, Tokyo 181-8588, Japan}
\affiliation{Yukawa Institute for Theoretical Physics, Kyoto University, Kyoto 606-8502, Japan}

\author{Kei Iida}
\affiliation{Department of Natural Science, Kochi University, 2-5-1 Akebono-cho, Kochi 780-8520, Japan
}

\author{Kazuhiro Oyamatsu}
\affiliation{Department of Human Informatics, Aichi Shukutoku University,
9 Katahira, Nagakute, Aichi 480-1197, Japan
}

\date{\today}

\begin{abstract}
Neutrons stars lighter than the Sun are basically composed of nuclear 
matter of density up to around twice normal nuclear density.  In our recent 
analyses, we showed that possible simultaneous observations of masses and 
radii of such neutron stars could constrain $\eta\equiv(K_0L^2)^{1/3}$, a 
combination of the incompressibility of symmetric nuclear matter $K_0$ and 
the density derivative of the nuclear symmetry energy $L$ that 
characterizes the theoretical mass-radius relation.  In this paper, we focus 
on the mass-radius constraint of the X-ray burster 4U 1724-307 given by 
Suleimanov et al. \cite{SPRW2011}.  We therefrom obtain the constraint 
that $\eta$ should be larger than around 130 MeV, which in turn leads to
$L$ larger than around 110, 98, 89, and 78 MeV for $K_0=180$, 230, 280, and 360
MeV.  Such a constraint on $L$ is more or less consistent with that obtained 
from the frequencies of quasi-periodic oscillations in giant flares 
observed in soft-gamma repeaters. 
\end{abstract}

%
\maketitle

{\it Introduction.}
Neutron stars, stellar remnants of supernova explosions at the end of 
massive stars, are considered to be composed of matter in extreme 
conditions, namely, ultra-high density and large neutron excess.  Since 
the temperature of the matter is generally very low compared with the 
typical neutron Fermi temperature, it is extremely difficult to examine 
the equilibrium properties of such dense cold matter in the laboratory, 
although highly energetic heavy-ion collisions could create hot dense matter as 
encountered in protoneutron stars.   Theoretically, on the other hand, the 
equation of state (EOS) for matter in neutron stars, hereafter referred to
as neutron star matter, remains to be determined, particularly above normal 
nuclear density, $\rho_0$.  Inversely, neutron stars could be a 
suitable laboratory to probe the properties of cold dense matter.  For 
example, observations of masses and radii of neutron stars would help us
to constrain the EOS of neutron star matter.  In fact, recent discoveries
of neutron stars with about two solar mass ($M_\odot$) play a role 
in ruling out various soft EOS models \citep{D2010,A2013}.
Furthermore, estimates of radiation radii of neutron stars have been made via 
observations of thermonuclear X-ray bursts and thermal spectra from 
low-mass X-ray binaries \citep{OBG2010,SLB2012,GSWR2013,LS2014}, which could 
also give us a significant constraint on the EOS.  Additionally, 
oscillation spectra radiated from a specific kind of neutron stars are 
another observable information to see stellar properties, such as masses, 
radii, the EOS, rotations, and magnetic fields (e.g., 
\cite{AK1996,AK1998,STM2001,SH2003,SKH2004,PA2011,SYMT2011,DGKK2013}).  This unique 
technique is known as neutron star asteroseismology.  Although 
observational evidences for neutron star oscillations are extremely 
limited, quasi-periodic oscillations discovered in the afterglow of giant 
flare phenomena observed from soft-gamma repeaters \citep{WS2006} are considered
to be strongly associated with oscillations of whatever portion of neutron
stars.  Through these observations, possible constraints on the stellar 
properties, particularly in the crustal region, are  
discussed \citep{SW2009,GNJL2011,S2011,SNIO2012,SNIO2013a,SNIO2013b,S2014}.

Since the details of neutron star structure obviously depend on the
still uncertain EOS of neutron star matter, they have yet to be 
clarified.  It is generally considered \citep{LaPr} that, under a 
liquid metallic ocean close to the surface, neutron-rich nuclei form a 
Coulomb lattice in a sea of electrons and, if any, dripped 
neutrons.  Because of the crystalline structure, the corresponding region 
is called a crust.  As the matter density increases up to a value 
close to $\rho_0$, it is considered that such nuclei begin to melt into 
uniform matter, which consists mainly of a core of the neutron star.
Furthermore, non-nucleonic components such as hyperons and quarks might appear 
for a still higher density region inside the core, depending on the  
model for neutron star matter \citep{NS}.  In addition to the possibility 
that such non-nucleonic components appear, it is also suggested that the 
uncertainty from three-neutron interactions in the EOS for pure neutron matter
comes into play for the same region \citep{GCR2012}. On the other hand, 
neutron star matter of density below about $2\rho_0$ is relatively 
easier to be constrained from terrestrial nuclear experiments.  This
is why we will focus particularly on low-mass neutron stars that have 
central density $\rho_c$ lower than $2\rho_0$.  We remark that we 
succeeded in constructing theoretical mass and radius formulae 
for such low-mass neutron stars, which are written as a function of 
$\rho_c$ and $\eta$, a combination of the EOS parameters that 
characterize the nuclear saturation properties \citep{SIOO2014}.

In this paper, we systematically examine the $\eta$ dependence of 
the mass-radius relation of low-mass neutron stars using more than 200 
phenomenological EOS models \citep{IO2014} that are constructed in such 
a way as to reproduce empirical masses and radii of stable nuclei.  Then, 
by comparing the obtained mass-radius relation with available neutron 
star observations, we give possible constraints on $\eta$.  In 
particular, for this purpose, we focus on constraints on the 
mass-radius relation of neutron stars that were derived
by Suleimanov et al. \cite{SPRW2011} from the observed cooling phases of the X-ray 
burster 4U 1724-307 located in the globular cluster Terzan 2 via 
different atmosphere models.  This is because unlike other studies 
to make a constraint on the mass-radius relation via the observed 
thermal emission from neutron stars, Suleimanov et al.\ obtained such 
constraints by using the whole cooling track and checking the consistency 
with the theoretical prediction of neutron star cooling evolution, which 
enables us to minimize the theoretical uncertainties in the atmosphere model 
during the burst phenomena \citep{SPW2011}.  As will be shown below, 
the resultant constraint on the density dependence of the 
symmetry energy is consistent with the known constraints from the 
quasi-periodic oscillations observed in giant flares of soft 
gamma repeaters \citep{SNIO2013b}.

\begin{figure}
\begin{center}
\includegraphics[scale=0.53]{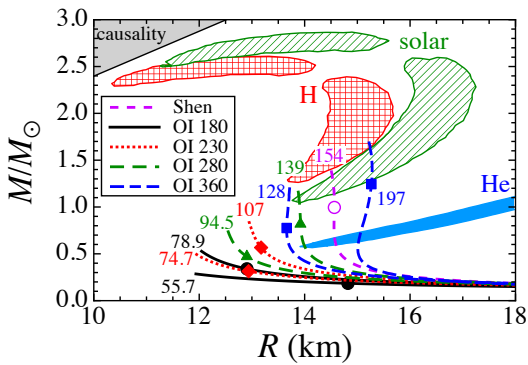} 
\end{center}
\caption{
(Color online) Allowed regions in the mass-radius relation 
obtained from the observation of the X-ray burster 4U 
1724-307 by \citet{SPRW2011}, where they adopted 
three different atmosphere models, i.e., pure hydrogen
(checkered region), pure helium (filled region), and the solar 
ratio of H/He with sub solar metal abundance $Z=0.3Z_\odot$ 
(shaded region).  On the other hand, the lines with marks 
denote the stellar models constructed from nine EOSs 
with different values of $\eta$ (attached numbers)
for $\rho_c\le 2.0\rho_0$, where each mark corresponds to
the mass and radius of a star with $\rho_c=1.5\rho_0$.
Additionally, the upper left region is ruled out by
the causality \citep{L2012}. 
}
\label{fig:MR}
\end{figure}

{\it EOS parameters.}
We begin with an expression for the EOS of uniform nuclear matter at 
zero temperature.  The bulk energy per nucleon, $w$, of this matter
can be generally expanded around the saturation point of symmetric nuclear 
matter as a function of the nucleon number density, $n_{\rm b}$, and neutron 
excess, $\alpha$, as \citep{L1981}
\begin{equation}
  w = w_0  + \frac{K_0}{18n_0^2}(n_{\rm b}-n_0)^2 + \left[S_0 
      + \frac{L}{3n_0}(n_{\rm b}-n_0)\right]\alpha^2,
  \label{eq:w}
\end{equation}
where $\alpha$ is defined as $\alpha=(n_{\rm n} - n_{\rm p})/n_{\rm b}$
with the neutron and proton number densities, 
$n_{\rm n}$ and $n_{\rm p}$.  That is, the case of $\alpha = 0$ corresponds 
to symmetric nuclear matter, while the case of $\alpha=1$ corresponds to pure 
neutron matter.  The parameters $w_0$, $n_0$, and $K_0$, which 
characterize this expansion, denote the saturation energy, saturation density,
and incompressibility of symmetric nuclear matter, respectively.  On the 
other hand, $S_0$ and $L$ are the parameters associated with the symmetry 
energy coefficient, i.e., $S_0$ is the symmetry energy coefficient at 
$n_{\rm b}=n_0$, and $L$ is the density dependence of the symmetry energy 
around $n_{\rm b}=n_0$.  Note that among the five parameters in Eq.\ 
(\ref{eq:w}), $w_0$, $n_0$, and $S_0$ can be relatively easier to 
constrain from empirical masses and radii of stable nuclei, while 
the remaining two parameters, $K_0$ and $L$, are more difficult to fix 
\citep{OI2003}.  Thus, we particularly focus this paper on 
the parameters $K_0$ and $L$.  We remark that many EOSs of nuclear 
matter have been proposed so far, which have various values of $K_0$ 
and $L$, while having reasonable values of $w_0$, $n_0$, and $S_0$ (e.g., 
\cite{OI2003,SIOO2014}).  Although it may well be difficult to   
precisely describe the mass-radius relation of neutron stars by taking $K_0$ 
and $L$ alone as free parameters, these two parameters are expected to mainly 
control the stiffness of the EOS of neutron-rich nuclear matter near $\rho_0$
and hence the structure of at least an outer part of neutron stars.  In fact,
we succeeded in finding a suitable combination of $K_0$ and $L$, namely, 
$\eta\equiv (K_0 L^2)^{1/3}$, that well characterizes the structure of
low-mass neutron stars \citep{SIOO2014} in the sense that the mass-radius 
relation changes smoothly with $\eta$ (see Fig.\ \ref{fig:MR}).

Now, in order to cover a wide range of $\eta$, we consider the 
phenomenological EOSs of neutron star matter based on the 
simplified version of the Thomas-Fermi method that allows for the 
bulk, gradient, and Coulomb energies \citep{OI2003,OI2007}.  
These EOS models were systematically obtained from the energy of
uniform nuclear matter, which, in the limit of $n_{\rm b}\to n_0$ 
and $\alpha\to 0$, reduces to Eq.\ (\ref{eq:w}) with various 
values of $y\equiv -K_0S_0/(3n_0L)$ and $K_0$.  In fact, the
most relevant values of $w_0$, $n_0$, and $S_0$ were determined
together with that of the gradient energy coefficient for given $y$ 
and $K_0$ by fitting masses and charge radii of 
stable nuclei obtained from the optimal nucleon distribution to the 
empirical ones \citep{O1993}.  We remark that $y$ corresponds to the 
gradient of the saturation line near $\alpha =0$.  Finally, the 
crustal EOS was obtained for various sets of $(L, K_0)$ 
\citep{OI2007} by extending the Thomas-Fermi method to several shapes 
of nuclei in a lattice within a Wigner-Seitz approximation \citep{O1993}. 
Hereafter, the resultant EOSs are referred to as the OI-EOSs.

\begin{figure}
\begin{center}
\includegraphics[scale=0.53]{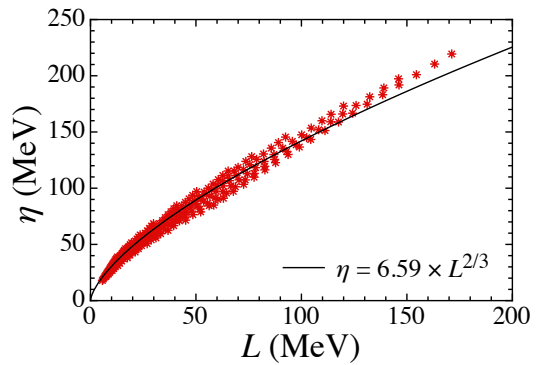} 
\end{center}
\caption{
(Color online) $\eta$ as a function of $L$.  The dots 
are taken from the 247 OI-EOSs, while the solid line denotes the 
fitting in a functional form of $L^{2/3}$. 
}
\label{fig:eta_L}
\end{figure}

\begin{figure}
\begin{center}
\includegraphics[scale=0.53]{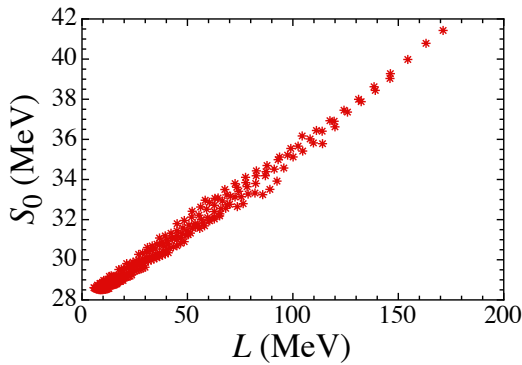} 
\end{center}
\caption{
(Color online) The parameter $S_0$ is plotted 
as a function of $L$ for the 247 OI-EOSs. 
}
\label{fig:LS0}
\end{figure}

The OI-EOSs adopted here have a range of $y< -200$ MeV fm$^3$ 
and $180\le K_0 \le 360$ MeV, which results in the range of $L$ as 
$0< L <180$ MeV.  Note that not only does such parameter range 
equally well reproduce empirical mass and radius data for 
stable nuclei, but also effectively covers even extreme cases 
\citep{OI2003}.  We also remark that according to 
comprehensive re-analysis of recent data on the giant monopole 
resonance energies, $K_0$ should be in the range of $250<K_0<315$ MeV 
\citep{SSM2014}, while the generally accepted value of $K_0$ is in the 
range of $K_0=230\pm40$ MeV \citep{KM2013}.  That is, systematic 
errors in experimentally determining $K_0$ are still likely to 
be large.  It is thus reasonable that the OI-EOSs used here
have 247 sets of $(y, K_0)$, i.e., the combination of 13 different 
values of $y$ ($y=-200$, $-220$, $-250$, $-300$, $-350$, $-400$, $-500$, 
$-600$, $-800$, $-1000$, $-1200$, $-1400$, and $-1800$ MeV fm$^3$) 
and 19 different values of $K_0$ ($K_0=180$, 190, 200, $\cdots$
360 MeV). For these OI-EOSs, the corresponding values of $\eta$ 
are calculated, which are shown in Fig.\ \ref{fig:eta_L} as a 
function of $L$.  From this figure, one can observe that the dependence 
of $\eta$ on $L$ is much stronger than that on $K_0$.  This is 
partly because uncertainties in $L$ are relatively large 
compared with those of $K_0$ and partly because the power of $L$ in
$\eta$ is larger than that of $K_0$.

Additionally, for comparison, we show $S_0$ and $L$ for the 247 OI-EOSs 
in Fig.\ \ref{fig:LS0}. A strong correlation between $S_0$ and $L$ 
was pointed out by \cite{OI2003}, and is consistent
with the values of $S_0$ and $L$ obtained via fitting to experimental data 
on nuclear masses and radii on $1\sigma$ level with the nuclear 
energy density functional for Skyrme type interaction \citep{EDF}. As
compared with this correlation, the correlation between $\eta$ and $L$
is equally strong.   Via simultaneous observations 
of masses and radii of low-mass neutron stars, therefore,
constraints on $\eta$ and thus $L$ would be available to some extent.
Finally, in Fig.\ \ref{fig:eta_L}, we also show the fitting to 
the data of the 247 OI-EOSs, i.e., 
$\eta=6.59 \left(\frac{L}{1~{\rm MeV}}\right)^{2/3}$
MeV, which corresponds to $K_0=286.8$ MeV.


{\it Constraints on $\eta$ and $L$.}
Following the finding of $\eta$, we here give a possible 
constraint on $\eta$ from observations of masses and radii of low-mass neutron 
stars.  Unfortunately, however, no firm observational evidence for
the presence of less-than-$1 M_\odot$ neutron stars is available.
Even more challenging is simultaneous mass and radius determination of 
low-mass neutron stars.  Thermonuclear X-ray bursts in low-mass X-ray 
binaries help to determine the masses and radii of the bursting 
neutron stars, although there exist many uncertainties both in theoretical 
models and in observations.  In fact, it is not straightforward to determine 
the exact moment when the luminosity reaches the Eddington limit at the star's
surface, if data for the photospheric radius expansion bursts are adopted
to determine the star's mass and radius.  Additionally, the color-correction 
factor defined as the ratio of the color temperature to the effective temperature of 
the source object is sensitive to the flux during the cooling tail as well as 
the model of neutron star atmospheres \citep{SPW2011}.  To minimize 
uncertainties in the theoretical models that determine the Eddington 
luminosities during the burst phenomena, Suleimanov et al.\ suggested using
information from the whole cooling track in the X-ray bursts, and succeeded 
in obtaining the constraint on the mass and radius of the X-ray burster 4U 1724-307 
located in the globular cluster Tarzan 2 by using various atmosphere models 
\citep{SPRW2011}.  In this paper, we adopt their results, which imply 
a relatively low-mass neutron star, to obtain a constraint on 
$\eta$ and $L$.

In particular, Suleimanov et al.\ adopted three atmosphere models with different 
chemical compositions, i.e., pure hydrogen, pure helium, and the solar H/He composition 
with sub solar metal abundance $Z=0.3Z_\odot$ appropriate for Terzan 2 \citep{O1997}. 
Then, assuming a flat distribution of the distance from the Earth between 5.3 
and 7.7 kpc with Gaussian tails of $1\sigma=0.6$ kpc, they obtained such
constraints in the mass-radius relation within 90\% confidence level as shown in 
Fig.\ \ref{fig:MR}, where the checkered, filled, and shaded regions correspond to the 
constants obtained with the atmosphere models composed of pure hydrogen, 
pure helium, and the solar H/He with $Z=0.3Z_\odot$, respectively.  In addition 
to their results, we show the region ruled out by the causality, which is 
given by $R < 2.824 GM/c^2$ \citep{L2012}.  From this figure, one can observe that 
the radius of the X-ray burster 4U 1724-307 should be relatively large if the 
star's mass has a canonical value of order $1.4M_\odot$.

In Fig.\ \ref{fig:MR}, we also plot the stellar models constructed with 
several sets of the OI-EOSs and the Shen EOS \citep{ShenEOS}, 
where the corresponding value of $\eta$ is written on each EOS. 
Here, we particularly focus on the stellar models for 
$\rho_c\le 2.0 \rho_0$ to avoid uncertainties in the EOS at high
density due to the possible appearance of non-nucleonic components 
and/or the profoundness of three-neutron interactions as mentioned above. 
In Fig.\ \ref{fig:MR}, therefore, the upper end of each line 
corresponds to the stellar model constructed with $\rho_c=2.0\rho_0$; 
for reference, we also show the stellar model for $\rho_c=1.5\rho_0$ 
by putting a mark on each line.

\begin{figure}
\begin{center}
\includegraphics[scale=0.53]{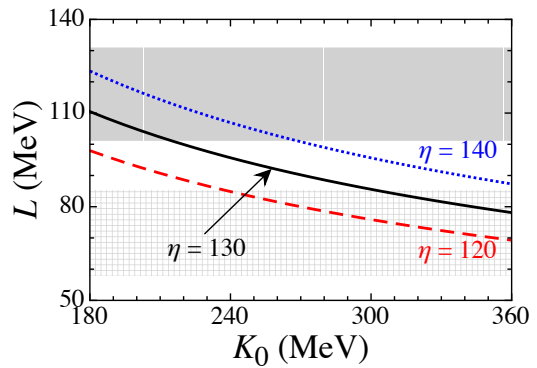} 
\end{center}
\caption{
(Color online) Nuclear matter EOS parameters constrained 
from $\eta \ge 130$ MeV, which correspond
to the region above the solid line.  For reference, 
$\eta= 120$ and 140 MeV are plotted with the 
dashed and dotted lines.  We also display the parameter 
space constrained from the observations of quasi-periodic
oscillations in giant flares with the filled and checkered 
regions \citep{SNIO2013b} (see text for details).
}
\label{fig:LK0}
\end{figure}

Now, assuming that the X-ray burster 4U 1724-307 has a canonical neutron 
star mass and that the EOS is universal in the sense that all 
neutron stars can be constructed with a single EOS, one can 
conclude from Fig.\ \ref{fig:MR} that $\eta$ is larger than 
$\sim130$ MeV.  Via $L = \sqrt{\eta^3/K_0}$, $\eta\gtrsim 130$ MeV
leads to $L \gtrsim 110$ MeV for $K_0=180$ MeV, $L\gtrsim 98$ MeV for 
$K_0=230$ MeV, $L\gtrsim 89$ MeV for $K_0=280$ MeV, and $L\gtrsim 78$ MeV 
for $K_0=360$ MeV.  One can more clearly see the allowed region in the 
parameter space in Fig.\ \ref{fig:LK0}, i.e., the region above the solid 
line, where we show the lines for $\eta=120$ MeV (dashed line) and 140 
MeV (dotted line) for reference.  We remark that the density at the
core-crust boundary strongly depends on the value of $L$ \citep{OI2007}, 
which is a crucial property to determine the crust mass and 
moment of inertia.  Combining Fig.\ 5 in Ref. \cite{OI2007} with 
the constraint on $L$ obtained here from the neutron star 
observations, the nucleon number density at the crust basis is expected 
to be around 0.07 fm$^{-3}$.

Recently, a lower limit of observed neutron star radii has been
additionally suggested from another object.  That is, the neutron star 
radius in the low-mass X-ray binary 4U 1608-52 is predicted 
from the hard-state burst occurring during the low-accretion rate
to be larger than 13 km, if the neutron star mass is in the 
range of 1.2-2.4$M_\odot$ \citep{Poutanen2014}.  This suggestion 
also indicates a large value of $\eta$, i.e., $\eta\gtrsim100$ 
MeV, which covers the constraint obtained from the X-ray burster 
4U 1724-307.  Thus, $\eta$ is predicted to be larger than around 
130 MeV from both of the astronomical observations.

We conclude this section by noting that observations of 
neutron star oscillations could also tell us the properties 
of neutron star matter.  In fact, the gravitational waves 
emitted from oscillating neutron stars could provide 
a possible way to see the neutron star properties, although 
they have not yet been observed directly.  Another
possibility is the detection of electromagnetic waves associated 
with neutron star oscillations \cite{SK2013,SK2014}. 
Fortunately, quasi-periodic 
oscillations have been discovered in the afterglow of giant 
flares observed from soft-gamma repeaters \citep{WS2006}. 
If such oscillations result from crustal torsional 
oscillations, a fairly strong constraint on $L$ can be obtained
by comparing the observed frequencies of quasi-periodic 
oscillations with theoretical predictions of the 
eigen-frequencies of the torsional modes
\citep{SNIO2012,SNIO2013a,SNIO2013b}.  In this way, 
we obtained two possible constraints on $L$. 
One is $101.1\le L\le 131.0$ MeV, which explains all the
observed frequencies lower than 100 Hz in terms of the 
crustal torsional oscillations, while the other is 
$58.0\le L\le 85.3$ MeV if the second lowest frequency 
observed in SGR 1806--20 would be excited by a different mechanism 
from the crustal oscillations.  These constraints 
on $L$ are also shown in Fig.\ \ref{fig:LK0} by the filled 
region for the former constraint and by the checkered 
region for the latter one. As can be seen from 
this figure, the former constraint on $L$ is more 
consistent with the constraint from $\eta\gtrsim 130$ MeV 
than the latter one. 
On the other hand, we remark that most of the terrestrial nuclear
experiments suggest somewhat lower values of $L$ \cite{Tsang2012},
although there still exists large uncertainty in $L$ \cite{Newton2013}.
We also remark that the EOS of pure neutron matter calculated within
the chiral effective field theory favors smaller values of $L$ (see \cite{HS2014}
and references therein).
Anyway, $\eta$ (and $L$) could be significantly smaller than our constraint,
if constraints on $M$ and $R$ of several neutron stars
(e.g., \cite{OBG2010,SLB2012,GSWR2013,LS2014}), other than the
ones adopted in the present analysis, are taken for granted.

{\it Conclusion.}
We apply the classification of the EOS of neutron star matter 
in terms of $\eta$ as developed in Ref. \cite{SIOO2014} to 
the mass-radius relation constrained from observations of the 
X-ray burster 4U 1724-307 and remark on a possible constraint
on $\eta$, which gives important information on the density
dependence of the symmetry energy.  In order to obtain a 
better constraint on $\eta$, it would be significant to 
expect simultaneous mass and radius determination of 
low-mass neutron stars from the neutron star interior 
composition explorer (NICER) by NASA and/or the large observatory 
for X-ray timing (LOFT) by ESA via observations of 
pulse profiles from hot rotating neutron stars \citep{NICER}.

\acknowledgments
H.S. is grateful to V. Suleimanov and J. Poutanen for providing us
with their data in preparing Fig.\ \ref{fig:MR}. This work was 
supported in part by Grants-in-Aid for Scientific Research on 
Innovative Areas through No.\ 24105001 and No.\ 24105008 provided by MEXT, 
by Grant-in-Aid for Young Scientists (B) through No.\ 26800133 provided by JSPS,
by the Yukawa International Program for Quark-hadron Sciences, and 
by the Grant-in-Aid for the global COE program ``The Next Generation of Physics, 
Spun from Universality and Emergence" from MEXT.



\end{document}